\documentclass[aip]{revtex4-2}

\draft 

\usepackage{graphicx} 
\usepackage{multirow}
\usepackage{longtable}
\usepackage{xcolor}
\usepackage{amsmath}
\usepackage[flushleft]{threeparttable}
\definecolor{lightblue}{rgb}{0.68, 0.85, 0.9}
\definecolor{blue(munsell)}{rgb}{0.0, 0.5, 0.69}

\graphicspath{{./}{figures/}}

\begin{document}


\title{The Infrared Absorption Spectrum of Phenylacetylene and its Deuterated Isotopologue in the Mid- to Far-IR} 



\author{Vincent J. Esposito}
\email[]{Vincent.J.Esposito@nasa.gov}
\affiliation{NASA Ames Research Center, MS 245-6, Moffett Field, CA 94035, USA}

\author{Piero Ferrari}
\affiliation{Radboud University, Institute for Molecules and Materials, HFML-FELIX, 6525 ED Nijmegen, The Netherlands}

\author{Wybren Jan Buma}
\affiliation{Radboud University, Institute for Molecules and Materials, HFML-FELIX, 6525 ED Nijmegen, The Netherlands}
\affiliation{Van't Hoff Institute for Molecular Sciences, University of Amsterdam, 1098 XH Amsterdam, The Netherlands}

\author{Ryan C. Fortenberry}
\affiliation{Department of Chemistry \& Biochemistry, University of Mississippi, University, MS 38677-1848, USA}

\author{Christiaan Boersma}
\affiliation{NASA Ames Research Center, MS 245-6, Moffett Field, CA 94035, USA}

\author{Alessandra Candian}
\affiliation{Anton Pannekoek Institute for Astronomy, University of Amsterdam, 1098 XH Amsterdam, The Netherlands}

\author{Alexander G. G. M. Tielens}
\affiliation{Leiden Observatory, Leiden University, 2333 CA Leiden, The Netherlands}
\affiliation{Astronomy Department, University of Maryland, College Park, MD 20742-2421, USA}


\date{\today}

\begin{abstract}
Anharmonicity strongly influences the absorption and emission spectra of polycyclic aromatic hydrocarbon (PAH) molecules. Here, \textcolor{black}{IR-UV} ion-dip spectroscopy experiments together with detailed anharmonic computations reveal the presence of fundamental, overtone, as well as 2- and 3-quanta combination band transitions in the far- and mid-infrared absorption spectrum of phenylacetylene and its singly deuterated isotopologue. Strong absorption features in the 400-900~cm$^{\rm -1}$ range originate from CH(D) in-plane and out-of-plane wags and bends, as well as bending motions including the C$\equiv$C and CH bonds of the acetylene substituent and the aromatic ring. For phenylacetylene, every absorption feature is assigned either directly or indirectly to a single or multiple vibrational mode(s). The measured spectrum is dense, broad, and structureless in many regions but well characterized by computations. Upon deuteration, large isotopic shifts are observed. At frequencies above 1500~cm$^{\rm -1}$ for d$_1$--phenylacetylene, a one-to-one match is seen when comparing computations and experiment with all features assigned to combination bands and overtones. The C$\equiv$C stretch observed in phenylacetylene is not observed in d$_1$--phenylacetylene due to a computed 40-fold drop in intensity. Overall, a careful treatment of anharmonicity that includes 2- and 3-quanta modes is found to be crucial to understand the rich details of the infrared spectrum of phenylacetylene. Based on these \textcolor{black}{results}, it can be expected that such an all-inclusive anharmonic treatment will also be key for unraveling the infrared spectra of PAHs in general.
\end{abstract}


\maketitle 

\section{Introduction}
\label{intro}
PAHs are a class of molecules composed of fused hexagonal rings of sp$^2$-hybridized carbon atoms with hydrogens decorating the periphery, which are known to be ubiquitous in space based on their infrared emission features detected throughout many interstellar regions, i.e., the \textcolor{black}{aromatic infrared bands (AIBs)}.\cite{tielens2008interstellar,boersma_jwst_2023} Recently, experimental and computational studies of the infrared absorption spectra of various PAHs have revealed large anharmonic effects. Previous attempts to account for anharmonicity included the use of harmonic scaling factors.\cite{laury2012vibrational,langhoff_theoretical_1996,bauschlicher_calculation_1997,bauschlicher_nasa_2010,bauschlicher_nasa_2018,pirali_high-resolution_2009,behlen1981fluorescence,cane1996high,bakker_infrared_2011,roithova_electronic_2019,wiersma_gas-phase_2022} 
\textcolor{black}{However, scaling factors do not account for important higher-order transitions such as combination bands and overtones, nor do they provide a good treatment of systems with a shallow minimum in their PESs.}\cite{lemmens2023water}

In cases where anharmonicity is strong, scaling factors are therefore not able to account for -and accurately predict- infrared absorption spectra, in particular for regions that can only be explained when anharmonic transitions such as combination bands are correctly taken into account.\cite{esposito_anharmonic_2023,esposito_anharmonicity_2023,mackie_anharmonic_2015,mackie_anharmonic_2016,mackie_anharmonic_2018,mackie_fully_2018,mackie_anharmonicity_2022,banhatti_formation_2022,pirali_high-resolution_2009,mulas_anharmonic_2018} Nowadays, with the launch of the James Webb Space Telescope (JWST), very sensitive and high-resolution infrared spectral data of the PAH bands have and will continue to be delivered,\cite{chown2023pdrs4all} calling for the need of a deeper understanding of anharmonic effects on the vibrational spectra of PAHs, and how to take them into account into astronomical models. To do so requires a synergy between accurate computational modeling and sensitive laboratory spectroscopy experiments. 


In the recent past dedicated gas-phase experiments have been performed for a number of PAHs, and these indeed reveal that for these compounds anharmonicity is essential to come to a proper interpretation of their infrared absorption spectra.  For example, \textcolor{black}{IR-UV} ion-dip molecular beam experiments have been used to record infrared absorption spectra of neutral naphthalene, anthracene, tetracene and pentacene, showing the appearance of many combination bands \textcolor{black}{in the 1600-2000~cm$^{\rm -1}$ region}.\cite{maltseva_high-resolution_2015,mackie_anharmonic_2015,mackie_anharmonic_2016,maltseva_high-resolution_2016,maltseva_high-resolution_2018,lemmens2019anharmonicity} Similarly, molecular beam experiments on the larger PAHs coronene, peropyrene, ovalene, and hexa(peri)benzocoronene, have shown the presence of many combination bands and overtones in \textcolor{black}{both} the 1600–2000~cm$^{\rm -1}$ and 2950–3150~cm$^{\rm -1}$ regions, whereas at lower frequencies (100 and 1000~cm$^{\rm -1}$) fundamental transitions dominate.\cite{lemmens2021infrared} 
Recently, using an infrared table-top laser system to cover the 3~\textmu m spectral region, \textcolor{black}{our} research group performed ion-dip experiments on phenylacetylene and d$_1$--phenylacetylene
(the singly deuterated isotopologue of phenylacetylene with deuteration taking place at the acetylene hydrogen),\cite{esposito_anharmonicity_2023} representing an ideal model system to assess the impact of anharmonic effects on PAHs, in particular focusing on the possibility that higher-order quanta modes are a significant contributor to their infrared spectra. Remarkably, a major influence of 2- and 3-quanta modes arising from aromatic and acetylenic C-H and C-D stretching bands is observed.\cite{esposito2023anharmonicity} Notably, phenylacetylene was recently detected in the dark molecular cloud TMC-1 via its pure rotational transitions,\cite{loru2023detection} making the detailed understanding of its infrared spectrum a pressing need.   

In this work, the previous study is extended to the much larger 100--2300~cm$^{\rm -1}$ spectral range using an infrared free electron laser, thereby covering the far- and mid-infrared regions. Detailed anharmonic quantum chemical calculations are utilized to reveal the key role played by higher-order modes in shaping the infrared spectra.   

\section{Methods}
\label{methods}

\subsection{Experimental}
Molecular beams of phenylacetylene and d$_1$--phenylacetylene are formed by placing the liquid compounds in a reservoir kept at room temperature that is seeded with a stream of Ar gas injected at a backing pressure of 4 bar. Using a Series 9 pulsed valve from General Valve, the Ar-phenylacetylene mixtures are expended into vacuum, forming a cold and directional molecular beam that is collimated by a 2~mm skimmer. The pulsed valve is operated at 10~Hz. After collimation, the molecular beam enters the ionization region of a reflectron time-of-flight mass spectrometer (R.M. Jordan D-850) equipped with a 40~mm dual microchannel plate (MCP) detector (Jordan Co. C-726) that has a mass resolution of $m/\Delta m$ = 2200 at 100~amu. \textcolor{black}{In previous studies on the same apparatus with similarly-sized molecular systems, molecular dynamic simulations were used to compute infrared spectra at different temperatures, finding good agreement with a vibrational temperature of 50 K.\cite{lemmens2023water} Similarly, previous estimations of the molecular beam temperature place its value in the range from 10 to 50 K.\cite{yatsyna_infrared_2016,jaeqx_gas-phase_2014} Values much higher than this range would lead to visible hot bands in REMPI scans, which is not the case in the spectra in this study.}

To measure mass spectra, phenylacetylene is ionized using a (1+1) Resonance Enhanced Two-Photon Ionization (R2PI) scheme. UV excitation is performed by the second harmonic light of a dye laser (Radiant Dye) pumped by the third harmonic of a Nd:YAG laser (InnoLas SpitLight1200) and operating on Coumarin 153 in ethanol. The laser system is tuned to the vibrationless origin of the S$_1$ $\leftarrow$ S$_0$ electronic transition of each species occurring at 35875.7 and 35888.6~cm$^{\rm -1}$ for phenylacetylene and d$_1$--phenylacetylene, respectively.\cite{esposito2023anharmonicity}   


 Infrared spectra are measured by ion-dip spectroscopy. For this, the laser light of the free electron laser FELIX is aligned counter-propagating with the molecular beam as described in Ref.\cite{bakels2020gas}. Upon resonant vibrational excitation, depopulation of the vibrational ground-state leads to a reduction of the ion signal. Running the molecular beam instrument at 10~Hz and operating FELIX at 5~Hz then allows the consecutive measurement of mass spectra with and without FELIX excitation and thereby the recording of IR absorption spectra. Importantly, FELIX is timed 300~{\textmu}s prior to the R2PI ionization, ensuring that vibrational excitation occurs when the molecules are in their neutral state. FELIX is tuned from 100 to 2300~cm$^{\rm -1}$ in steps of 1~ cm$^{\rm -1}$. 

\subsection{Computational}

Computation of the optimized geometry; harmonic normal modes; and quadratic, cubic, and quartic normal coordinate force constants (quartic force field; QFF) is performed at the B3LYP\cite{becke_densityfunctional_1993}/N07D\cite{barone_development_2008} level of theory utilizing the Gaussian 16 software package.\cite{m._j._frisch_g._w._trucks_h._b._schlegel_g._e._scuseria_m._a._robb_j._r._cheeseman_g._scalmani_v._barone_g._a._petersson_h._nakatsuji_et_al._gaussian_2016} \textcolor{black}{The computations were done using the verytight optimization criteria and a custom integration grid consisting of 200 radial shells and 974 angular points per shell (compared to 99 radial shells and 590 angular points per shell in the default UltraFine grid)} The N07D basis set is based on the 6-31G(d) basis set and contains additional diffuse and polarization functions that have been shown to increase accuracy in the anharmonic computations of large aromatic systems such as PAHs.\cite{barone_fully_2014} A QFF is a truncated Taylor series expansion of the potential surface surrounding the equilibrium geometry, following the formula:

\begin{equation}
  \begin{split}
     V = & \frac{1}{2}\sum_{i,j}^{3N} {\biggl (\frac{\partial^2 V}{\partial X_i\partial X_j}\biggl)X_iX_j} \\
     + & \frac{1}{6}\sum_{i,j,k}^{3N} {\biggl (\frac{\partial^3 V}{\partial X_i\partial X_j\partial X_k}\biggl)X_iX_jX_k} \\
     + & \frac{1}{24}\sum_{i,j,k,l}^{3N} {\biggl (\frac{\partial^4 V}{\partial X_i\partial X_j\partial X_k\partial X_l}\biggl)X_iX_jX_kX_l} \\
  \end{split}
\end{equation}
\textcolor{black}{The QFF is computed in normal mode coordinates via a linear relationship to produce a Cartesian coordinate QFF. Additionally, semi-diagonal quartic terms are employed as is the default in Gaussian16.\cite{barone_anharmonic_2005}} 

From here, two different methodologies are implemented via 2$^{\rm nd}$ order vibrational perturbation theory (VPT2).\cite{fortenberry_chapter_2019,franke_how_2021,watson_vibrational_1977,fortenberry_lee_vibrational_2022} Using the built-in anharmonic frequency code within Gaussian 16, VPT2 is used to compute the anharmonic frequencies including up to three-quanta transitions. This allows for the inclusion of multi-mode combination bands that are imperative for an accurate analysis of the recorded experimental absorption data.

Alternatively, \textcolor{black}{to properly treat the redistribution of intensity as a result of accidental resonances between vibrational states}, a locally modified version of SPECTRO\cite{gaw_spectro_1991} is utilized to compute the anharmonic vibrational frequencies of transitions involving changes up to two vibrational quanta. SPECTRO has the advantage of using resonance polyads in the anharmonic computations.\cite{martin_accurate_1997,martin_anharmonic_1995} When two \textcolor{black}{vibrational states} are close in frequency and produce a near-singularity in the VPT2 treatment, they are removed from the standard perturbation treatment and included in a resonance polyad matrix. The matrix allows for the treatment of resonance effects as well as normal modes that simultaneously participate in multiple resonances. Additionally, the polyad matrix allows for the distribution of intensities between coupled vibration\textcolor{black}{al states. Vibrational modes with frequencies below 300~cm$^{\rm -1}$ are excluded from the SPECTRO VPT2 calculation}. This method has been described previously in greater detail.\cite{maltseva_high-resolution_2015,mackie_anharmonic_2015,mackie_anharmonic_2016,mackie_anharmonic_2018,mackie_anharmonicity_2022,esposito_anharmonic_2023,esposito_anharmonicity_2023}

The resultant vibrational absorption stick spectrum is convoluted with a Gaussian line shape having a full-width at half-maximum (FWHM) of 0.5\% of the \textcolor{black}{computational frequency} as to match the bandwidth of FELIX.

\section{Results and Discussion}
\label{results}

\subsection{Phenylacetylene Absorption Spectrum}
\label{phen_abs_sec}

\begin{figure}
  \includegraphics[width=\textwidth]{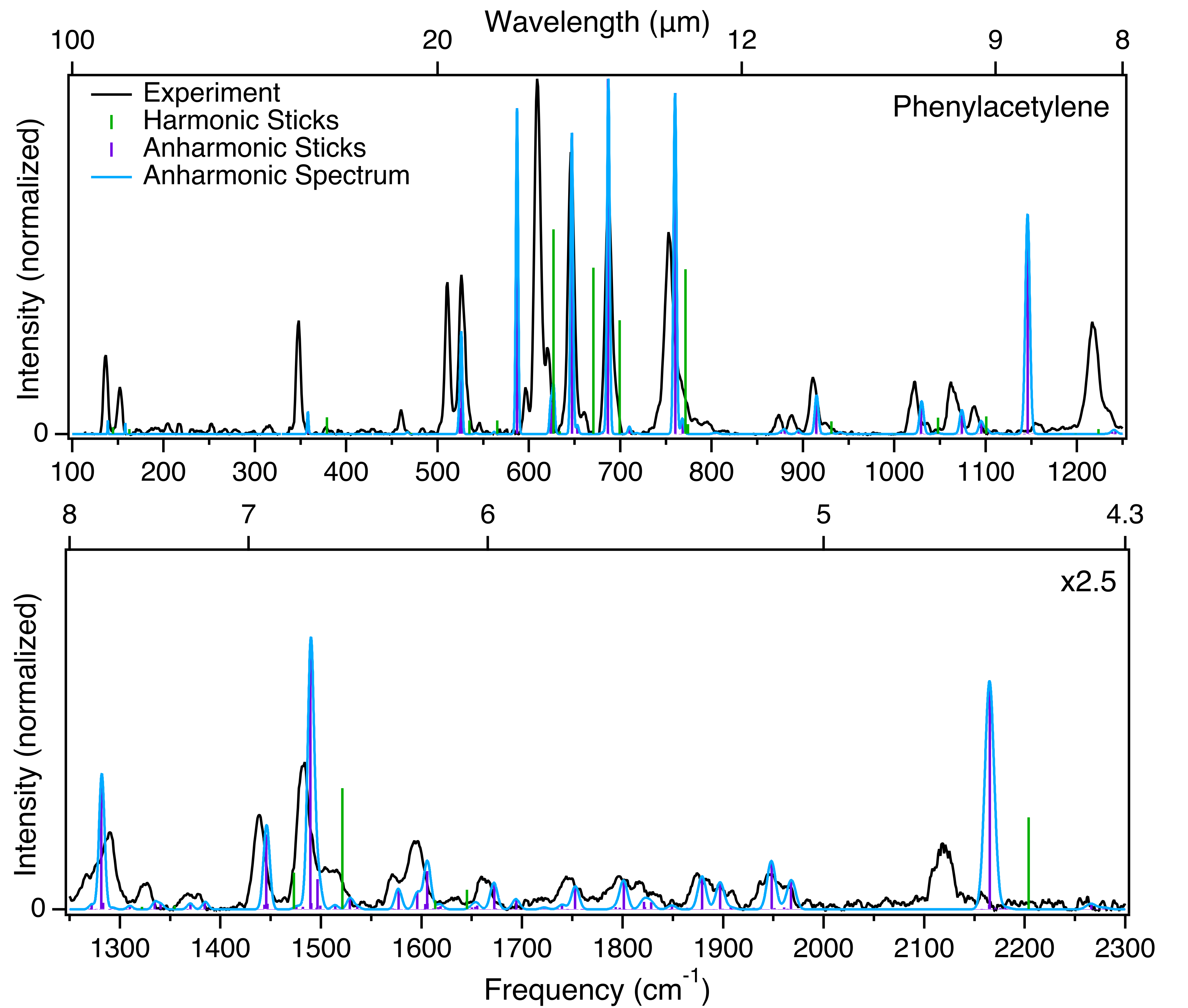}
  \caption{Experimental (black)\textcolor{black}{, harmonic stick, and broadened} anharmonic (blue; stick spectrum in purple) absorption spectrum of phenylacetylene. The computed anharmonic spectrum in purple represents the stick spectrum convoluted with a Gaussian function with a FWHM of 0.5\% of the \textcolor{black}{computational} frequency to reproduce the experimental linewidth of FELIX. The intensity in the lower panel has been multiplied by a factor of 2.5 to show details of the weaker features in this region.}
  \label{phen_abs}
 \end{figure}

Figure~\ref{phen_abs} presents the experimental infrared absorption spectrum (black) of phenylacetylene in the 100-2300~cm$^{\rm -1}$ frequency range, \textcolor{black}{ the harmonic stick spectrum (green)}, the 3-quanta anharmonic stick spectrum (purple), and the artificially broadened anharmonic spectrum (blue) based on this stick spectrum. The experimental band center, FWHM, and relative intensity along with the pertinent associated computational anharmonic mode labeling, frequency, and intensity of the pertinent bands are presented in Table~\ref{phen_table}. The full list of \textcolor{black}{computational frequencies and experimental transition assignments} are provided in Table~S1 of the Supplementary Information (SI). The experimental transitions are determined via a least-squares fit using a Gaussian line shape. Overall, approximately 40 features can be identified from the experimental data above a signal-to-noise ratio of 2. Because of the resolution of the experiment, the many overlapping bands leads to unstructured, broad shapes and shoulders. Studying the computational spectrum leads to a deeper understanding as to why many of the features appear so broad and structure-less. For example, in the experimental spectrum the fit detected two neighboring features at 753.1 and 764.8~cm$^{\rm -1}$ within the broad feature centered at 753.1~cm$^{\rm -1}$, which exhibits a shoulder on the high-frequency side. Using the anharmonic computations, these two features can be reliably assigned to $\nu_{19}$ (out-of-plane aromatic CH wag) and $\nu_{12}$ (in-plane aromatic ring breathing mode), respectively, clearly explaining the finer details of the 753.1~cm$^{\rm -1}$ feature.

 \setlength{\tabcolsep}{4pt}
 \begin{table}
 \begin{threeparttable}
 \scriptsize
 \caption{Phenylacetylene mode number, anharmonic computational frequency (cm$^{\rm -1}$), intrinsic intensity (km mol$^{\rm -1}$), and relative intensity; and the experimental band centers (cm$^{\rm -1}$), FWHM (cm$^{\rm -1}$), and relative intensities.}
 \begin{tabular}{p{2cm}p{2cm}p{2.5cm}p{1.7cm}p{1.5cm}p{1.5cm}p{1.cm}}
 \hline
 \hline
 Mode & Frequency (cm$^{\rm -1}$) & Intrinsic   Intensity (km mol$^{\rm -1}$) & Comp.   Rel. Intensity\tnote{a} & Exp.   Band Centers (cm$^{\rm -1}$) & FWHM   Gaussian (cm$^{\rm -1}$) & Exp.   Rel. Int. \\ \hline
  $\nu_{21}$ & 586.8 & 38.910 & 0.917 & 609.4 & 5.9 & 1.000 \\
  $\nu_{13}$+$\nu_{36}$ & 626.7 & 5.733 & 0.135 & 620.6 & 6.1 & 0.267 \\
  $\nu_{23}$+$\nu_{24}$+$\nu_{36}$ & 653.0 & 1.078 & 0.025 & 659.2 & 6.8 & 0.049 \\
  $\nu_{19}$ & 760.3 & 40.727 & 0.960 & 753.1 & 8.4 & 0.588 \\
  $\nu_{12}$ & 768.3 & 1.856 & 0.044 & 764.8 & 13.1 & 0.155 \\
  $\nu_{21}$+$\nu_{24}$+$\nu_{36}$ & 876.2 & 0.277 & 0.007 & 873.3 & 5.8 & 0.056 \\
  $\nu_{10}$ & 1029.5 & 3.902 & 0.092 & 1021.8 & 7.8 & 0.138 \\
  $\nu_{32}$ & 1073.9 & 2.855 & 0.067 & 1064.0 & 10.7 & 0.155 \\
  $\nu_{13}$+$\nu_{34}$ & 1095.3 & 1.536 & 0.036 & 1087.6 & 7.1 & 0.073 \\
  2$\nu_{21}$ & 1146.0 & 26.230 & 0.618 & 1217.6 & 11.6 & 0.267 \\
  $\nu_{7}$ & 1489.6 & 12.538 & 0.296 & 1481.9 & 10.2 & 0.162 \\
  \hline
  2$\nu_{21}$+$\nu_{33}$ & 1793.3 & 0.109 & 0.003 & \multirow{2}{*}{1795.9} & \multirow{2}{*}{18.6} & \multirow{2}{*}{0.037} \\
  $\nu_{14}$+$\nu_{15}$ & 1801.4 & 1.375 & 0.032 &  &  &  \\
  \hline
  $\nu_{14}$+$\nu_{18}$ & 1879.0 & 1.579 & 0.037 & \multirow{3}{*}{\begin{tabular}[c]{@{}c@{}}1875.4\\    \\ 1896.3\end{tabular}} & \multirow{3}{*}{\begin{tabular}[c]{@{}c@{}}15.5\\    \\ 21.5\end{tabular}} & \multirow{3}{*}{\begin{tabular}[c]{@{}c@{}}0.037\\    \\ 0.022\end{tabular}} \\
 $\nu_{17}$+$\nu_{18}$ & 1897.2 & 1.291 & 0.030 &  &  &  \\
 3$\nu_{33}$ & 1907.5 & 0.092 & 0.002 &  &  &  \\
 $\nu_{5}$ &2165.3&	10.637	&0.251&	2119.0&	18.0&	0.078\\
 \hline
 \hline
 \end{tabular}
 \begin{tablenotes}
   \item[a] Computational relative intensities are in reference to the transition at 686.5~cm$^{\rm -1}$ which is not listed here but can be found in Table S1 of the Supplementary Information  
  \end{tablenotes}
  \end{threeparttable}
 \label{phen_table}
 \end{table}

Across the entirety of the spectrum (Figure~\ref{phen_abs}), every experimental feature is assigned to either a single or multiple vibrational modes. Interestingly, the two features at 609.4 and 1217.6~cm$^{\rm -1}$ stand out as having poor agreement with the computations. The data in Table~\ref{phen_table} show that the feature centered at 609.4~cm$^{\rm -1}$ is assigned to $\nu_{21}$, which is the out-of-plane \textcolor{black}{(OOP)} acetylene CH wag that causes a large change in the dipole moment and leads to high intensity. The difference between the experiment and computed position for this mode is 22.6~cm$^{\rm -1}$, larger than expected for a fundamental mode at this level of theory, especially considering that the surrounding features are well-described by the computations (e.g., the feature at 620.6~cm$^{\rm -1}$ is assigned to the $\nu_{13}$~+~$\nu_{36}$ combination band computed to be at 626.7~cm$^{\rm -1}$). Anharmonic computations of \textcolor{black}{OOP} bending motions of aromatic systems consisting of sp$^2$ hybridized carbons have been previously reported to exhibit unexpected behavior when using DFT methods in combination with some basis sets,\cite{lee_unsolved_2021}. \textcolor{black}{In this study, different DFT and correlated methods with varied basis sets were tested and all suffered the same issues with the OOP bending mode, and in fact, the B3LYP/N07D method performed the best against experiment. Therefore, we believe that this is indeed the cause of the issue seen here. To confirm that there are no issues with basis set size, the harmonic vibrational spectrum of phenylacetylene calculated with B3LYP/N07D compares well to those calculated with B3LYP/6-311++G(d,p), agreeing to 16~cm$^{\rm -1}$ in the CH stretch region and approximately 4~cm$^{\rm -1}$ for lower frequency modes. The in-plane acetylene CH wag is slightly more sensitive to basis set size; however, a difference of approximately 3\% will not impact the results in this study.} Additionally, phenylacetylene has been reported to exhibit large couplings between various \textcolor{black}{vibrational states of} the acetylene substituent.\cite{esposito2023anharmonicity} The initial error in the potential energy surface computation for $\nu_{21}$ is compounded when moving to higher-order transitions along with the general decrease in accuracy when computing the frequency of overtone transitions \textcolor{black}{(see Ref. \cite{fortenberry_excited_2015} for more details).} This leads to a larger error in the frequency for the feature centered at 1217.6~cm$^{\rm -1}$, which arises from the first overtone of $\nu_{21}$ ($2\nu_{21}$). The other feature that has poor agreement between experiment and computation is the feature at 2119.0~cm$^{\rm -1}$, assigned to $\nu_{5}$ (2165.3~cm$^{\rm -1}$), the acetylene C$\equiv$C stretch. It is unclear at the moment why there is such a large difference between experiment and theory for this feature.


Moving to higher frequencies, each feature from 1250-2000~cm$^{\rm -1}$ is assigned to one or more underlying modes. Due to the congested nature of this region, some features are assigned to multiple nearby modes that cannot be separated effectively. \textcolor{black}{The harmonic stick spectrum (green) in Figure~\ref{phen_abs}} illustrates how important it is to include anharmonicity in the computations of the vibrational spectrum of PAHs. In the 1250-2000~cm$^{\rm -1}$ frequency range, only seven fundamental frequencies (green) exist, making it impossible to completely assign the complicated spectrum of phenylacetylene. However, when anharmonicity is included, hundreds of relatively intense overtone and combination bands present themselves that agree quite nicely with the experimental features. Furthermore, the anharmonic intensities show better agreement with the experimental intensities than the double harmonic intensities do. 

\begin{figure}
\includegraphics[width=\textwidth]{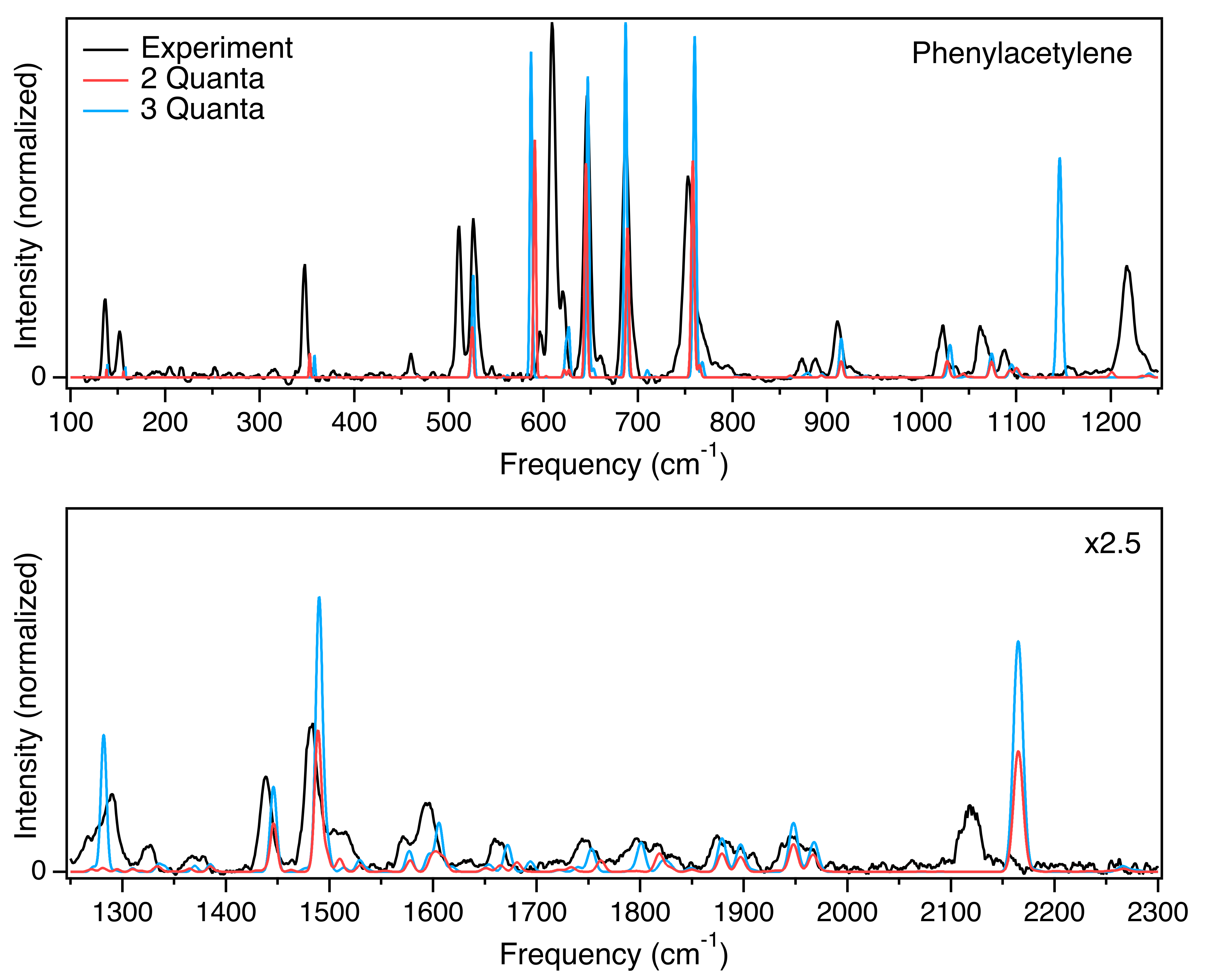}
\caption{Experimental (black), 3-quanta anharmonic (blue), and 2-quanta anharmonic (red) absorption spectrum of phenylacetylene. The computed 3-quanta anharmonic spectrum shown in blue is convoluted with a Gaussian line shape with a linewidth equal to 0.5\% of the \textcolor{black}{computational} frequency, reproducing the experimental linewidth of FELIX, while the 2-quanta anharmonic spectrum shown in red has a constant 1.0~cm$^{\rm -1}$ FWHM (for visualization purposes). The intensity in the lower panel has been multiplied by a factor of 2.5 to show details of the weaker features in this region.}
\label{phen_2v3}
\end{figure}

Figure~\ref{phen_2v3} illustrates the 2-quanta (red) and 3-quanta (blue) vibrational modes computed for phenylacetylene. The importance of including the 3-quanta modes cannot be overstated, as can be seen from the figure and is highlighted by Table~\ref{phen_table}. At least four 3-quanta modes (combination bands involving three fundamental vibrations or one fundamental mode and one overtone, as well as a second overtone mode) are assigned to features in the spectrum that would have been missed if the higher-order modes would not have been included in the computations. For example, around 1800~cm$^{\rm -1}$, the experiment reveals a double feature that can only be explained by including 3-quanta modes. Of note is that the 3-quanta computations are not performed with polyad matrices and therefore do not benefit from the proper treatment of intensity sharing. The less accurate intensities can be seen in the systematically larger intensity for almost all modes in the 3-quanta computations versus those of the 2-quanta in Figure~\ref{phen_2v3}. \textcolor{black}{A glaring example of this is the disappearance of the 2$\nu_{21}$ feature in the 3-quanta computation. As the result of many strong resonance coupling interactions, the intrinsic intensity of this mode is redistributed over various nearby features.} Not \textcolor{black}{utilizing a polyad matrix} will also affect the frequencies of the bands in the 3-quanta computations, although to a lesser degree. These issues will be addressed in subsequent studies, when the capability to compute 3-quanta modes is developed further.

\subsection{d$_1$--Phenylacetylene Absorption Spectrum}

\begin{figure}
\includegraphics[width=\textwidth]{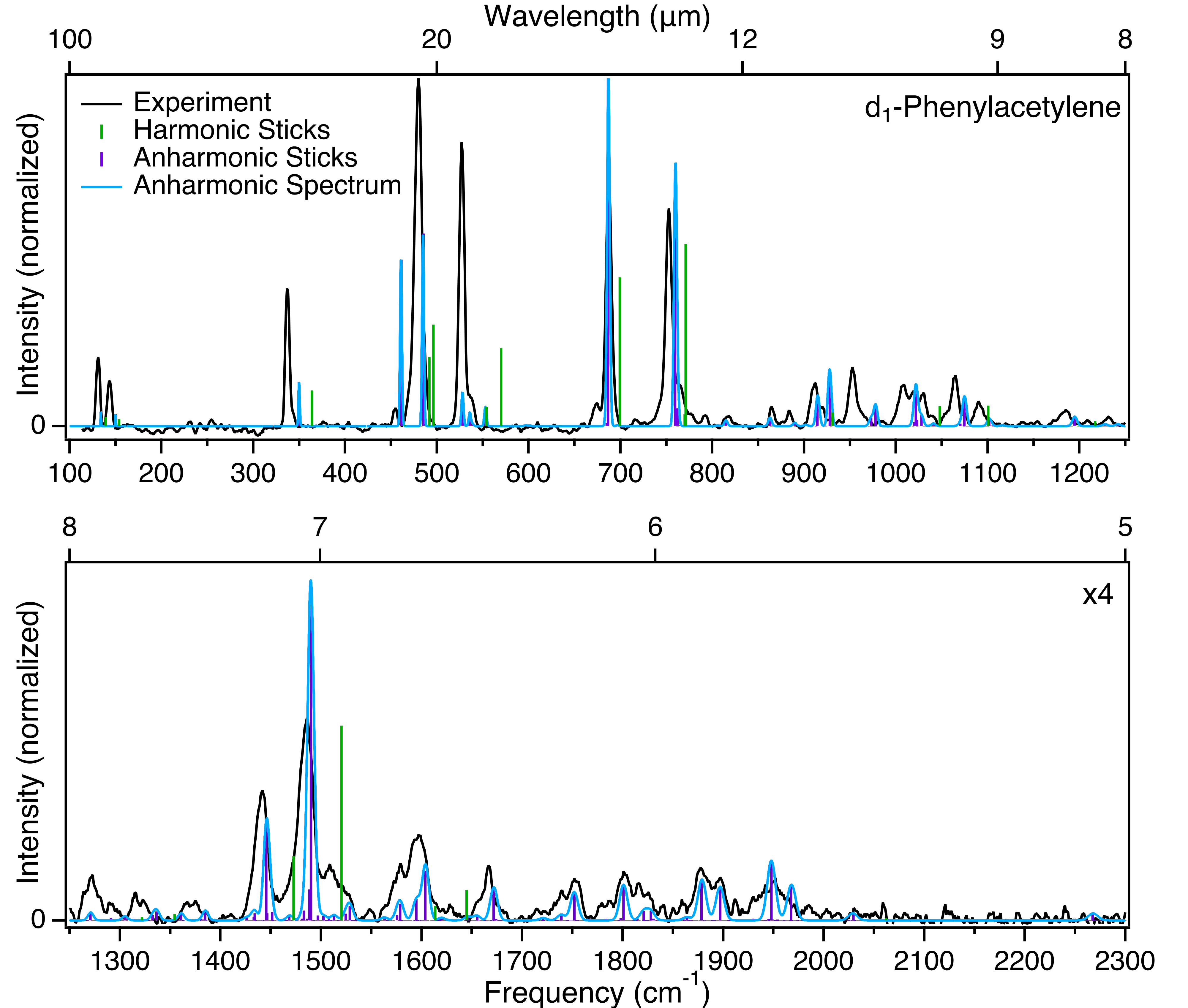}
\caption{Experimental (black)\textcolor{black}{, harmonic stick, and} anharmonic (blue; stick spectrum in purple) absorption spectrum of d$_1$--phenylacetylene. The computed anharmonic spectrum in blue represents the stick spectrum convolved with a Gaussian linewidth equal to 0.5\% of the \textcolor{black}{computational} frequency to reproduce the experimental linewidth of FELIX. The intensity in the lower panel has been multiplied by a factor of 4 to show details of the weaker features in this region.}
\label{Dphen_abs}
\end{figure}

Figure~\ref{Dphen_abs} presents the absorption spectrum of d$_1$--phenylacetylene in the range of 100-2300~cm$^{\rm -1}$ with the experiment shown in black,\textcolor{black}{ the harmonic stick spectrum in green,} the anharmonic 3-quanta stick spectrum in purple, and the anharmonic stick spectrum broadened with a Gaussian lineshape (FWHM of 0.5\% of the \textcolor{black}{computational} frequency) in blue. Table~\ref{Dphen_table} lists computational and experimental frequencies of prominent bands. The full list of experimental and computational frequencies is provided in Table~S2 of the SI. In the experimental spectrum bands are observed throughout the spectral range with mostly sharp, well-defined peaks below 800~cm$^{\rm -1}$ and broad, structure-less features above. The sole physical difference between phenylacetylene and d$_1$--phenylacetylene is a single deuteration, but this leads to major spectroscopic changes. Many spectral shifts originate from modes that involve motion of the deuterium atom, and adjacent modes (such as the acetylene C$\equiv$C stretch) are also impacted. Here, theory provides insights into what drives these spectral changes. 

\setlength{\tabcolsep}{4pt}
 \begin{table}
  \begin{threeparttable}
 \scriptsize
 \caption{d$_1$-Phenylacetylene mode number, anharmonic computational frequency (cm$^{\rm -1}$), intrinsic intensity (km mol$^{\rm -1}$), and relative intensity; and experimental band centers (cm$^{\rm -1}$), FWHM (cm$^{\rm -1}$), and relative intensities.}
 \begin{tabular}{p{2cm}p{2cm}p{1.2cm}p{1.5cm}p{1cm}p{1.5cm}p{1.2cm}}
 \hline
 \hline
 Mode & Frequency (cm$^{\rm -1}$) & Intrinsic   Intensity (km mol$^{\rm -1}$) & Comp.   Rel. Intensity & Exp.   Band Centers (cm$^{\rm -1}$) & FWHM   Gaussian (cm$^{\rm -1}$) & Exp.   Rel. Int. \\ \hline
$\nu_{22}$ & 461.5 & 21.666 & 0.478 & 475.7 & 3.8 & 0.731 \\
$\nu_{13}$ & 461.7 & 0.245 & 0.005 & 454.3 & 2.0 & 0.079 \\
$\nu_{35}$ & 485.4 & 25.055 & 0.553 & 480.6 & 5.3 & 1.000 \\
$\nu_{21}$ & 528.1 & 4.407 & 0.097 & 527.2 & 4.8 & 0.828  \\
$\nu_{16}+\nu_{24}$ & 535.9 & 1.785 & 0.039 & 537.2 & 5.8 & 0.082 \\
$\nu_{34}$ & 539.1 & 0.325 & 0.007 & & &  \\
\hline
$\nu_{21}+$2$\nu_{22}$ & 1426.2 & 0.119 & 0.003 & \multirow{6}{*}{1440.8} & \multirow{6}{*}{11.9} & \multirow{6}{*}{0.104} \\
$\nu_{14}$+$\nu_{22}$ & 1433.3 & 0.132 & 0.003 &  &  &  \\
$\nu_{18}$+$\nu_{21}$ & 1434.2 & 0.272 & 0.006 &  &  &  \\
$\nu_{28}$ & 1445.5 & 3.588 & 0.079 &  &  &  \\
$\nu_{19}$+$\nu_{20}$ & 1446.6 & 0.270 & 0.006 &  &  &  \\
$\nu_{17}$+$\nu_{22}$ & 1451.4 & 0.321 & 0.007 &  &  &  \\
\hline
2$\nu_{21}$+$\nu_{22}$ & 1468.6 & 0.190 & 0.004 & \multirow{9}{*}{\begin{tabular}[c]{@{}l@{}}1485.1\\    \\ 1508.8\\    \\ 1521.9\end{tabular}} & \multirow{9}{*}{\begin{tabular}[c]{@{}l@{}}14.7\\    \\ 9.5\\    \\ 10.8\end{tabular}} & \multirow{9}{*}{\begin{tabular}[c]{@{}l@{}}0.160\\    \\ 0.043\\    \\ 0.036\end{tabular}} \\
$\nu_{13}$+$\nu_{34}$+$\nu_{35}$ & 1483.1 & 0.380 & 0.008 &  &  &  \\
$\nu_{10}$+$\nu_{13}$ & 1488.7 & 1.185 & 0.026 &  &  &  \\
$\nu_{7}$ & 1490.2 & 11.941 & 0.264 &  &  &  \\
$\nu_{8}$+2$\nu_{36}$ & 1496.8 & 0.190 & 0.004 &  &  &  \\
$\nu_{13}$+$\nu_{20}$+$\nu_{23}$ & 1502.5 & 0.169 & 0.004 &  &  &  \\
2$\nu_{19}$ & 1512.9 & 0.209 & 0.005 &  &  &  \\
2$\nu_{12}$ & 1524.6 & 0.257 & 0.006 &  &  &  \\
\hline
$\nu_{15}$+$\nu_{20}$ & 1528.8 & 0.538 & 0.012 &  &  &  \\
$\nu_{10}$+$\nu_{34}$ & 1563.2 & 0.116 & 0.003 & \multirow{6}{*}{\begin{tabular}[c]{@{}l@{}}1576.6\\    \\ 1596.8\end{tabular}} & \multirow{6}{*}{\begin{tabular}[c]{@{}l@{}}12.0\\    \\ 15.9\end{tabular}} & \multirow{6}{*}{\begin{tabular}[c]{@{}l@{}}0.029\\    \\ 0.060\end{tabular}} \\
$\nu_{20}$+$\nu_{23}$+$\nu_{34}$ & 1575.8 & 0.194 & 0.004 &  &  &  \\
$\nu_{27}$ & 1578.8 & 0.642 & 0.014 &  &  &  \\
$\nu_{15}$+$\nu_{19}$ & 1594.9 & 0.812 & 0.018 &  &  &  \\
$\nu_{18}$+$\nu_{20}$ & 1603.9 & 0.240 & 0.005 &  &  &  \\
$\nu_{6}$ & 1604.0 & 1.902 & 0.042 &  &  &  \\
\hline
$\nu_{14}$+$\nu_{18}$ & 1878.6 & 1.578 & 0.035 & 1879.3 & 13.0 & 0.031 \\
$\nu_{17}$+$\nu_{18}$ & 1897.2 & 1.297 & 0.029 & 1896.8 & 8.5 & 0.025 \\
$\nu_{5}$ & 2029.5 & 0.277 & 0.006 &  &  & \\
\hline 
\hline 
\end{tabular}
\begin{tablenotes}
   \item[a] Computational relative intensities are in reference to the transition at 686.6~cm$^{\rm -1}$ which is not listed here but can be found in Table S2 of the Supplementary Information  
  \end{tablenotes}
  \end{threeparttable}
\label{Dphen_table}
\end{table}

There is generally good agreement between experiment and computation across the entire spectral range with a few notable outliers. In the range from 450-560~cm$^{\rm -1}$, there are four features present in the experimental spectrum while theory predicts six bands. The three largest peaks in the experimental spectrum occur at 475.7, 480.6 and 527.2~cm$^{\rm -1}$, with the peaks at 475.7 and 480.6~cm$^{\rm -1}$ overlapping (Figure~S1 shows an expanded view of this feature). An unambiguous assignment is difficult to make here, but the strongest peak at 480.6~cm$^{\rm -1}$ most likely originates from $\nu_{35}$, which is the in-plane acetylene CD wag. $\nu_{22}$, the OOP acetylene CD wag computed at 461.5~cm$^{\rm -1}$, is tentatively assigned to the experimental feature at 475.7~cm$^{\rm -1}$. This mode is analogous to $\nu_{21}$ in the standard isotopologue. Deuteration of the acetylene hydrogen leads to an experimental isotopic shift of 128.8~cm$^{\rm -1}$ for the OOP acetylene CH wag mode. Assigning the feature at 527.2~cm$^{\rm -1}$ is more difficult due to three modes computed near the central peak position and the intensity mismatch. The carrier of this peak is probably $\nu_{21}$, the OOP acetylene C$\equiv$C bend predicted at 528.1~cm$^{\rm -1}$. After the VPT2 treatment, the computed intensity of $\nu_{21}$ drops from 18.229~km mol$^{\rm -1}$ (0.524 relative to $\nu_{20}$) to 4.408~km mol$^{\rm -1}$ (0.097 relative to $\nu_{20}$). As mentioned previously, OOP bending modes are sometimes not well described by DFT/VPT2, and intensities in general are difficult to calculate. Therefore, better agreement is found when considering the double harmonic intensities. Additionally, the lack of intensity sharing in these calculations contributes to the intensity mismatch for all of the features mentioned here. The other two small peaks predicted at 535.9 ($\nu_{16}$+$\nu_{24}$) and 539.1 ($\nu_{34}$)~cm$^{\rm -1}$ add to the intensity of the central peak, and lead as well to the appearance of the shoulder on the high frequency side. 

Moving to higher frequencies, there is somewhat less agreement between experiment and computation in the 900-1100~cm$^{\rm -1}$ range. However, 10 prominent peaks are seen in both spectra, \textcolor{black}{providing} the ability to understand which modes are involved without directly assigning individual peaks. There is a mixture of transitions in this region characterized by fundamental, overtone, and 2- and 3-quanta combination bands contributing to features observed here; 2$\nu_{22}$ reports the largest predicted intensity (see SI Table~S2). Only three fundamental transitions have appreciable intensity in this region \textcolor{black}{(harmonic sticks in Figure~\ref{Dphen_abs})}, indicating and reinforcing the strong impact of anharmonicity on phenylacetylene \cite{esposito2023anharmonicity} and the need for anharmonic computations to properly analyze and understand its spectrum. 

The broad feature centered at 1440.8~cm$^{\rm -1}$ with a FWHM of 11.9~cm$^{\rm -1}$ mainly originates from the $\nu_{28}$ aromatic CH in-plane bend fundamental transition. Underlying 2- and 3-quanta combination bands account for the remaining intensity that leads to the broadness of the feature. At higher frequencies, the experimental fit identifies three overlapping peaks at 1485.1, 1508.8, and 1521.9~cm$^{\rm -1}$ that make up the double peak feature around 1500~cm$^{\rm -1}$. The two main modes here are the $\nu_{7}$ aromatic CH in-plane bend fundamental as well as the $\nu_{10}$+$\nu_{13}$ combination band transition. Other higher-order overtone and combination band transitions contribute to the broad intensity and double peak structure.

From 1550~cm$^{\rm -1}$ upwards, there is an almost one-to-one match of experimental and computational features, all of which are higher-order modes (see also SI Figure~S\textcolor{black}{2}). For example, $\nu_{18}$+$\nu_{19}$ is the combination band responsible for the experimental peak observed at 1666.4~cm$^{\rm -1}$, while $\nu_{14}$+$\nu_{18}$ (1878.6~cm$^{\rm -1}$) and $\nu_{17}$+$\nu_{18}$ (1897.2~cm$^{\rm -1}$) are responsible for the band observed experimentally at 1879.3 and 1896.8~cm$^{\rm -1}$, respectively. The acetylene C$\equiv$C stretch ($\nu_{5}$, 2029.5~cm$^{\rm -1}$) is not observed in the spectrum of d$_1$--phenylacetylene due to a drop in intensity upon deuteration, contrasting with the fairly strong feature observed in the standard isotopologue. 

\section{Conclusions}

Anharmonicity is intrinsically required for the accurate characterization of the vibrational spectrum of polycyclic aromatic hydrocarbons,\cite{esposito_anharmonicity_2023} a principle that is further supported by the results presented in the present study that covers frequencies from the mid- to far-infrared. For both phenylacetylene and d$_1$--phenylacetylene, direct and indirect assignments are only possible by including anharmonic terms in the vibrational potential energy surface calculation. The mid- to far-IR spectra of these molecules are full of higher-order transitions such as overtones and combination bands, including both 2- and 3-quanta combination bands. So far, 3-quanta bands have not received much attention. The present study demonstrates, however, that inclusion of such transitions is necessary to come to a proper understanding of these spectra. Additionally, the use of polyad matrices to allow for intensity sharing is integral for the correct treatment of anharmonic intensities in PAHs. Combined with the previous publication\cite{esposito_anharmonicity_2023} of the absorption spectra of phenylacetylene and d$_1$--phenylacetylene in the CH (CD) stretch fundamental region, the full absorption spectrum up to 3400 and 3100~cm$^{\rm -1}$, respectively, excluding the 2300-2500~cm$^{\rm -1}$ region that is outside of the laser frequency capabilities of FELIX, is now complete and included in Figure~S\textcolor{black}{3}. 

Overall, very good agreement is present between experiment and theory for both phenylacetylene and d$_1$--phenylacetylene with an average difference\textcolor{black}{ in frequency} of 1.2 and 1.0\%, respectively, for the directly assigned features. The low-frequency rocking and bending motions at approximately 500~cm$^{\rm -1}$ and below have some of the largest differences\textcolor{black}{ in frequency} with an average difference of 2.5 and 2.7\%. These modes are notoriously difficult to characterize with computational methods, and all of them deviate in a similar manner. Even so, an assignment is still easy to make for all features due to the sparsity of features in this region. As described in Section~\ref{phen_abs_sec}, $\nu_{21}$ and 2$\nu_{21}$ in phenylacetylene show the largest deviation\textcolor{black}{ in frequency} between theory and experiment (3.8 and 6.1\%) with the inaccuracy compounded for the overtone transition. This same deviation is not observed in d$_1$--phenylacetylene. The average difference on the order of 1\% for these molecules is larger than that found for unsubstituted\cite{mackie_anharmonic_2016} and methyl-substituted\cite{mackie_anharmonic_2018} PAHs. This is not so surprising given the strong resonances present in phenylacetylene \cite{esposito_anharmonicity_2023} and the known difficulty in computing vibrational spectra of acetylene molecules. \cite{fortenberry_lee_vibrational_2022} 

Large isotopic frequency shifts are reported here for various modes that involve the acetylene hydrogen. Many of these modes either lose or gain intensity based on deuteration, causing major changes to the resultant absorption spectrum, particularly in the 500-900~cm$^{\rm -1}$ (20-11.1~{\textmu}m) range. Further experiments on isotopically labeled PAHs are needed, and such  \textcolor{black}{absorption} experiments are currently underway in \textcolor{black}{our} laboratory. Even with the insights provided herein, there remain quite a number of unanswered questions regarding how deuteration impacts the emission of PAHs, especially with regards to the strong infrared bands observed in the 1-20~{\textmu}m range.\cite{boersma_jwst_2023,tielens2008interstellar,peeters_rich_2002} \textcolor{black}{Very little PAH, and their deuterated isotopologues in particular, emission data exist in the literature. A recent emission experiment on phenylacetylene and d$_1$--phenylacetylene carried out by Lacinbala et al.,\cite{lacinbala_aromatic_2022} and a follow-up computational study by our group\cite{esposito_anharmonicity_2023} look into the role that anharmonicity plays in the emission of substituted aromatic molecules. Outside of this recent example, there exists a dearth of experimental emission data for this wavelength region, requiring }further experimental and computational investigation to understand how anharmonicity and the isotopic frequency and intensity shifts impact the emission spectrum of standard and functionalized PAHs, molecules that are strongly tied to infrared observations pertinent to astronomical observation during the age of JWST and its possible successors. 


%
%

%
\section{Supplementary Material}
The data that support the findings of this study are available in the supplemental information and from the corresponding author upon reasonable request. 

\begin{acknowledgments}
V.J.E. acknowledges an appointment to the NASA Postdoctoral Program at NASA Ames Research Center, administered by the Oak Ridge Associated Universities through a contract with NASA. C.B. is grateful for an appointment at NASA Ames Research Center through the San Jos\'{e} State University Research Foundation (80NSSC22M0107). V.J.E, C.B., and R.F acknowledge support from the Internal Scientist Funding Model (ISFM) Laboratory Astrophysics Directed Work Package at NASA Ames. Computer time from the Pleiades cluster of the NASA Advanced Supercomputer (NAS) is gratefully acknowledged. R.C.F. acknowledges support from NASA grant NNH22ZHA004C and the Mississippi Center for Supercomputing Research supported in part from NSF Grant OIA-1757220. Studies of interstellar PAHs at Leiden Observatory are supported through a NWO Spinoza grant. The HFML-FELIX Laboratory is supported by the project CALIPSOplus under the Grant Agreement 730872 from the EU Framework Programme for Research and Innovation HORIZON 2020. We gratefully acknowledge the Nederlandse Organisatie voor Wetenschappelijk Onderzoek (NWO) and thank the FELIX staff. 
\end{acknowledgments}

\bibliographystyle{unsrt}
\bibliography{my_library}

\end{document}